\PassOptionsToPackage{colorlinks=true,pdfstartview=FitV,linkcolor=blue,citecolor=blue,urlcolor=blue,breaklinks=true}{hyperref}

\documentclass[10pt,aps,prd,reprint,showpacs,groupedaddress,eqsecnum,notitlepage,nofootinbib,superscriptaddress]{revtex4-2}

\usepackage{amsmath,amssymb,amsfonts,bm}
\usepackage{slashed}
\usepackage[T1]{fontenc}
\usepackage[utf8]{inputenc} 

\usepackage{graphicx}
\usepackage[caption=false]{subfig} 

\usepackage{tensor}
\usepackage{siunitx}
\usepackage{tikz}
\usepackage{accents}
\usepackage{scalerel}

\usepackage{hyperref} 

\usepackage{orcidlink}

\begin{document}

\title{Holonomic quantum computation on graphene from Atiyah-Singer index theorem}
\author{G. Q. Garcia\,\orcidlink{0000-0003-3562-0317}}
\email{gqgarcia99@gmail.com (Corresponding author)}
\affiliation{Centro de Ci\^encias, Tecnologia e Sa\'ude, Universidade Estadual da Para\'iba, 58233-000, Araruna, PB, Brazil.}
\author{M. Dantas}
\email{mirleide@df.ufcg.edu.br}
\affiliation{Unidade Acad\^emica de F\'isica, Universidade Federal de Campina Grande, 58429-140, Campina Grande, PB, Brazil. }
\author{A. M. de M. Carvalho\,\orcidlink{0009-0006-3540-0364}}
\email{alexandre@fis.ufal.br}
\affiliation{Instituto de F\'{\i}sica, Universidade Federal de Alagoas, 57072-970, Macei\'o,  AL, Brazil.}
\author{C. Furtado\,\orcidlink{0000-0002-3455-4285}}
\email{furtado@fisica.ufpb.br}
\affiliation{Departamento de F\'isica, Universidade Federal da Para\'iba, 58051-970, Jo\~ao Pessoa, PB, Brazil. }

\begin{abstract}
We investigate the emergence of geometric phases in graphene-based nanostructures through the lens of the Atiyah–Singer index theorem. By modeling low-energy quasiparticles in curved graphene geometries as Dirac fermions, we demonstrate that topological defects—arising from the insertion of pentagonal or heptagonal carbon rings—generate effective gauge fields that induce quantized Berry phases. We derive a compact expression for the geometric phase in terms of the genus and number of open boundaries of the structure, providing a topological classification of zero-energy modes. This framework enables a deeper understanding of quantum holonomies in graphene and their potential application in holonomic quantum computation. Our approach bridges discrete lattice models with continuum index theory, yielding insights that are both physically intuitive and experimentally accessible.
\end{abstract}

\maketitle

\section{Introduction}

The topological features of periodic lattices have garnered significant interest in condensed matter physics over the past decade. The discovery and synthesis of graphene by Geim and Novoselov~\cite{Science.306.666.2004} boosted the research into topological properties and laid the foundation for a new class of topological materials. The graphene lattice consists of a single layer of carbon atoms arranged in a two-dimensional honeycomb pattern. In the low-energy regime, its electronic excitations behave as massless Dirac fermions~\cite{PhysRevLett.69.1.1992, katsnelson2012graphene}, resulting in a linear energy-momentum dispersion relation. Because of this linear dispersion, the graphene lattice is regarded as a semi-metal with a zero bandgap, where its conduction band meets the valence band at the Fermi points. This pseudo-relativistic behavior leads to extremely high charge carrier mobility due to the absence of backscattering and effective mass~\cite{Nature.438.4233.2005}. This emergent Dirac-like dynamics is described by the effective Weyl Hamiltonian, which governs the behavior of low-energy quasiparticles in graphene, and its two triangular sublattices play the role of a pseudospin. This framework enables us to investigate a range of phenomena influenced by geometry and topology, including curvature effects and the presence of topological defects. Several graphene-based nanostructures have been developed, such as carbon nanotubes~\cite{ando2005theory, ando2009electronic}, fullerenes~\cite{gonzalez1993electronic, garcia2017geometric}, and nanocones~\cite{bueno2012landau, furtado2008geometric}, which can be created by introducing curvature or altering the lattice arrangement. These deformations introduce nontrivial topological features. 

A Weyl fermion feature strongly influenced by the system's geometry and topology is the geometric quantum phase. When an adiabatic evolution traces a closed path, the wavefunction acquires this phase related to the system's geometry~\cite{berry1984quantal}. Berry formulated the geometric quantum phase as a generalization of the phase obtained in the well-known Aharonov-Bohm effect. Especially in graphene, Berry phases are of utmost importance, encoding global properties of the system and offering a robust tool for characterizing the electronic structure of curved graphene. First, the probable presence of topological defects in graphene synthesis, through pentagon/heptagon rings, is responsible for the mixing of Fermi points in the spinor after adiabatic evolution~\cite{ContempPhys.50.2.2009}. As a consequence, a geometric quantum phase, in the special unitary group $SU(2)$, emerges to compensate for this blending. The geometric quantum phase is also used to classify the topological class of graphene. Breaking of fundamental symmetries in the graphene lattice~\cite{haldane1988model, kane2005quantum}, such as the inversion and time-reversal symmetries, leads to the emergence of a geometric phase, so that it is associated with a nonzero Chern number. Thus, graphene-based materials can be classified as topological insulators or not. Additionally, several graphene allotropes with different geometries have a geometric quantum phase related to the genus number~\cite{bakke2012kaluza, garcia2025landau, garcia2025rotation} by the index theorem. For the last case, we have focused our attention.    

In the literature, several index theorems have been developed, but in this work we focus on the Atiyah-Singer index theorem~\cite{AnnMath.87.3.1968}. This theorem allows for analyzing the spectral properties of certain elliptic differential operators, such as the Dirac operator, without requiring explicit diagonalization. At low energies, the effective Hamiltonian of graphene can be described by the Dirac operator. Therefore, the index theorem becomes a powerful tool for exploring geometrically modified structures derived from graphene, such as nanotubes and fullerenes. Specifically, we will use it to estimate the number of eigenstates with zero eigenvalue, known as zero modes. In systems with a symmetric energy spectrum, such as graphene, zero-energy states, also known as zero modes, play a crucial role in determining electrical conductivity, edge states, and the quantum Hall effect~\cite{katsnelson2012graphene, shen2017topological}. These modes are topologically protected, meaning that they do not vanish under smooth deformations of the structure. Their existence depends on the global geometric and topological features of the system, such as its genus and the number of open boundaries, rather than on the local details of the atomic lattice. This robustness makes zero modes a universal feature in graphene-based nanostructures with curvature and topological defects. Although graphene and its derivatives, like fullerenes, are made of discrete atomic lattices, it is common in the literature to model them as effective continuous surfaces for large-scale geometric and topological analysis. This approach is justified because the low-energy electronic excitations in these materials behave as Dirac fermions moving on a curved background. In this context, the index theorem remains applicable in the continuum approximation, and its predictions are supported by numerical and experimental studies.

In this paper, we analyze these geometric phases and explore how they enable the implementation of holonomic quantum computation in graphene-based nanostructures. Holonomic quantum gates emerge from adiabatic and cyclic evolutions that accumulate nontrivial holonomies in parameter space. The intrinsic curvature and topological defects in graphene provide a natural setting for such holonomies, making it a promising platform for geometric quantum logic. To this end, we develop a method to compute geometric phases arising in graphene nanostructures directly from their topological data, specifically the genus and the number of open boundaries, thus providing a purely topological classification of the quantum phases supported by these systems.

\section{Graphene Briefing}
\subsection{Tight-biding model}
Here, we present a mathematical description of the electronic states of graphene, where this material is a two-dimensional crystal composed of carbon atoms. The graphene lattice presents a hexagonal shape in which each carbon atom is localized on a hexagon vertex. The electronic properties of graphene can be described efficiently through the tight-binding model \cite{NuclPhysB.406.3.1993, RevModPhys.81.109.2009,Saito1998}. In this model, electrons tunneling from one site to its first neighbors of the lattice, while of course respecting the Pauli exclusion principle. Thus,
\begin{equation}\label{tight}
H=-J\left[\sum_{\left\langle i,j\right\rangle }a_i^\dagger a_j + h.c.\right],
\end{equation}
in equation $(\ref{tight})$, the term $J$ gives the intensity of the electron transition between neighbor sites. The indices $i$ and $j$ characterize the lattice sites, and here only the nearest neighbors will be considered. The terms $a_i^\dagger$ and $a_j$ are operators of the creation and annihilation of electrons at sites. The hexagonal bidimensional lattice of graphene is not a Bravais lattice. Thus, graphene is usually modeled by the superposition of two triangular lattices~\cite{PhysRevlett.56.26.1984}. From these considerations, the diagonalization of the graphene Hamiltonian~\cite{ContempPhys.50.2.2009} results in the following dispersion relation,
{\footnotesize
\begin{equation}\label{valores}
E(\textbf{k})=\pm J\sqrt{1+4\cos ^2\left( \dfrac{\sqrt{3}k_y}{2}\right)+4\cos\left(\frac{3k_x}{2}\right)\cos\left(\dfrac{\sqrt{3}k_y}{2}\right)}.
\end{equation}}
Note that when the dispersion relation is zero, the valence band is fully filled. The specific values of momentum for this condition connect the valence and conduction bands. These points are called Fermi points $K$ and $K'$. The existence of these two inequivalent Fermi points gives rise to a discrete degree of freedom known as the valley index. Electrons near the $K/K'$ points behave effectively as independent species analogous to spin, i.e., they have a pseudo-spin degree of freedom~\cite{Reich2002}. 

Expanding the Hamiltonian around the Fermi points and considering only terms of first-order expansion, we obtain~\cite{Peres2010},
\begin{equation}\label{dirac}
\begin{pmatrix} H_+ &  0 \\ 0 & H_- \end{pmatrix} \Psi(\textbf{r}) = \gamma \cdot \textbf{p} \, \Psi(\textbf{r}) = E \Psi(\textbf{r}),
\end{equation}
This equation reveals that the low-energy excitations in graphene behave as massless Dirac fermions in two dimensions. The spinor \( \Psi(\textbf{r}) \) encodes the amplitudes on the two sublattices $\mathcal{A/B}$ as well as the contributions of each Fermi point $K/K'$. The Dirac matrices \( \gamma \) operate both in the sublattice and in the valley space, and the momentum operator \( \textbf{p} = -i\hbar \nabla \) captures the long-wavelength behavior of the quasiparticles near the Dirac points. The emergence of this Dirac-like equation is a direct consequence of the symmetry and topology of the honeycomb lattice. It provides a powerful framework for understanding the high electronic mobility, the linear dispersion relation, and the semimetallic character of graphene. Moreover, this formulation sets the stage for exploring how external perturbations such as curvature, strain, magnetic fields, or topological defects can be incorporated through gauge fields, mass terms, or geometric couplings within the Dirac formalism.

\subsection{Introducing topological defects} 

The ability of graphene to be molded into different geometries is closely related to the presence of topological defects in its lattice. These topological defects are failures in the graphene lattice that occur during the synthesis process. They appear in the form of polygonal rings, differing from the hexagonal one, which adds curvature and/or torsion effects to the system. For instance, pentagon rings introduce a positive curvature, while heptagon rings introduce a negative one. The description of these topological defects was realized by Volterra through of ``cut and glue" processes~\cite{volterra1907equilibre}. To obtain a topological defect, it is necessary to remove or insert an angular sector of $m\pi/3$ and to glue its boundary. As a result, the sites of the same sublattice are connected. This reconnection alters the sublattice symmetry and can be interpreted as introducing an effective gauge field $A_\mu$ into the system. If a spinor is parallel transported around the resulting topological defect, it acquires a quantum phase. This situation is analogous to the Aharonov-Bohm effect, in which a wavefunction encircling a magnetic flux accumulates a phase proportional to that flux~\cite{Furtado2008, Lammert2000}. In this case, the effective gauge field simulates a fictitious magnetic flux generated by the lattice topology rather than a real magnetic field. Additionally, the effect of global geometry is captured by the spin connection $\Omega_\mu$, which appears in the covariant derivative and describes how spinors rotate along curved paths, as follows: 
\begin{equation}\label{GeoHolo}
    \oint_{C} \Omega_\mu\cdot d\textbf{r} = -\frac{\pi}{6} \sigma_z,
\end{equation}
where $\sigma_z$ is the Pauli matrix in the real space. The vector potential $A_\mu$ associated with the gauge field inserted to make the connection between the same two sites and to compensate for the mixing of Fermi points, is written as
\begin{equation}\label{abeliana0}
\oint_C \textbf{A} \cdot d \textbf{r}=\frac{\pi}{2}\tau_y.
\end{equation}
Here $\tau_y$ is the Pauli matrix in reciprocal space. Simplifying, we can reduce the above expression for each sub-Hamiltonian, and then
\begin{equation}\label{abeliana}
\oint_C \textbf{A} \cdot d \textbf{r}=\frac{\pi}{2},
\end{equation}
where $C$ is the pentagonal cell created and presented in $(\ref{abeliana})$ is the effective gauge flux \cite{EurPhysJST.148.127.2007}. And its corresponding magnetic field is defined by the expression $\textbf{B}=\nabla\times\textbf{A}$.

Therefore, a spinor transported around a curved region acquires a nontrivial holonomy, corresponding to a geometric phase that reflects the underlying curvature of the lattice. This is analogous to a Berry phase induced by geometry rather than external fields. Furthermore, the valley's degree of freedom can couple differently to the local geometry. Since curvature and topological defects break the inversion symmetry and induce localized strain fields, they can lift the degeneracy between valleys or introduce valley-dependent geometric phases. In these conditions, the Hamiltonian that best describes graphene in curved space is given by
\begin{equation}\label{curvatura}
H=\begin{bmatrix} -ie^{\mu}_{\ k}\sigma^k(\nabla_\mu - ieA_\mu) &  0 \\ 0 & ie^{\mu}_{\ k}\sigma^k(\nabla_\mu + ieA_\mu)\end{bmatrix}.
\end{equation}  
The zweibeins $e^{\mu}_{\ k}$ makes the covariant derivative $\nabla_\mu=\partial_\mu - i\Omega_\mu$, and $\Omega_\mu$ is the spin connection~\cite{Vozmediano2010}. 

\subsection{Atiyah-Singer index theorem on graphene} 

The presence of topological defects on the graphene lattice is responsible for the breaking of energy degeneracy, and, as a consequence, the emergence of a band gap and associated zero modes. One way to relate such zero modes to the geometric flux generated by the topological defect is through the Atiyah-Singer index theorem~\cite{AnnMath.87.3.1968}. The proof of such a theorem is described in Appendix~\ref {appxA}. In ref.~\cite{ContempPhys.50.2.2009}, Pachos obtained the Atiyah-Singer index theorem for several allotropes of graphene, with different geometries, making use of Euler's theorem. This theorem relates the structure of a polyhedron to topological properties. Thus, graphene allotropes can be described by a polyhedron that depends on the $V$ number of vertices, the $E$ edge number, and the $F$ faces number. Euler was the first to draw attention to the topological properties of polyhedra by a number known as the Euler characteristic. For a compact surface, the Euler characteristic can be written by
\begin{equation}\label{carac}
\chi=V-E+F=2(1-g),
\end{equation}
where the geometric characteristic of the polyhedron is related to the parameter $\textit{g}$, which represents the genus of the object under study. For open surfaces, we need to perform a normal cut in the compact surface, and then the Euler characteristic assumes the following form
\begin{equation}\label{carac2}
\chi=V-E+F=2(1-g) - N,
\end{equation}
 which $N$ is the number of open faces present in the polyhedron.

Therefore, if we consider that graphene allotrope is a polyhedron formed by heptagonal, hexagonal, and pentagonal rings. The amount of each of these polygonal rings will be represented, respectively, by $n_5$, $n_6$, and $n_7$. Thus, we can say that for polyhedra arising from graphene, the number of faces is given by $F=(n_5+n_6+n_7)$. In the case of vertices $V=(5n_5+6n_6+7n_7)/3$, because each ring has the number of vertices equal to the side number divided by $3$, because each vertex is linked with another three vertices. For instance, a pentagon has five sides and thus five vertices; however, each cell is shared by three different cells, so you need to divide by three. Similarly, the total number of edges is given by $E = (5n_5+6n_6+7n_7)/2$. Substituting these quantities into equation $(\ref{carac})$, Pachos obtained the Euler characteristic~\cite{ContempPhys.50.2.2009} for graphene. Considering the number of open faces of the structure as follows, we have that
\begin{equation}\label{general}   
\chi=\frac{n_5-n_7}{6}=2(1-g)-N.
\end{equation}
This equation indicates that the configuration of the {\it genus} on graphene depends on the relative number of pentagons and heptagons.

Using the well-known Stokes theorem, it is possible to relate this equation to the flow corresponding to each cell, pentagonal or heptagonal. Thus, we have obtained the total flow across the surface of the molecule, such that pentagons and heptagons contribute to flow in the opposite way, which one obtains.
\begin{equation}\label{diferenca}
\frac{1}{2\pi}\iint_{S}\textbf{B}\cdot d \textbf{S}=\frac{1}{2\pi}\sum_{n_5-n_7}\oint_{{C}_{c}}\textbf{A} \cdot d\textbf{r}=\frac{1}{2}(n_5-n_7).
\end{equation}
Thus, the problem comes down to find the difference between the number of pentagons and heptagons existing in the molecule. However, this quantity has been obtained in equation $(\ref{general})$ by the Euler characteristic and therefore,
\begin{equation}\label{quase}
\frac{1}{2\pi}\iint_{S}\textbf{B}\cdot d \textbf{S}=3(1-g)-\frac{3}{2}N.
\end{equation}
The total number of eigenstates with zero energy is the sum of contributions of each $H_\pm$ sub-operator of the Hamiltonian in eq.~\eqref{curvatura}. Considering the results obtained in appendix~\ref{appxA}, especially the equations $(\ref{final})$ and $(\ref{quase})$, we can say that the index of the Hamiltonian, which describes the molecules derived from graphene, is given by
\begin{equation}\label{index}
\textit{index}(H)=\frac{1}{\pi}\iint_{S}\textbf{B}\cdot d \textbf{S}=6(1-g)-3N.
\end{equation}
The equation $(\ref{index})$ relates the minimum number of zero modes of geometric variants of graphene to its topological characteristics, taking into account the degeneracy of the system. In particular, it applies very well to fullerenes. It is easy to see that for this molecule $index(H)=6$, i.e., it has at least six zero modes. This result agrees with numerical calculations, which showed that the fullerene has two triplet states with energies very close to zero \cite{NuclPhysB.406.3.1993}. With the help of equation $(\ref{index})$, it is possible to observe geometric phases in graphene molecules arising from their topological characteristics.

\section{Atiyah-Singer Geometric Phase on Graphene}

The geometric phases are global quantum effects acquired by a system undergoing adiabatic evolution along a closed path in the parameter space. These phases play a central role in holonomic quantum computation, where logical operations are implemented through unitary transformations that depend solely on the geometry of the path~\cite{PhysLettA.264.2.1999}. Since the initial proposals, the idea has been widely developed and realized on experimental platforms~\cite{PhysRevLett.110.19.2013}, including condensed matter systems such as graphene~\cite{EurophysLett.87.3.2009}. In this context, a quantum logic gate can be understood as a unitary transformation of a quantum state:
\begin{equation}
{\left | \psi \right \rangle}_f = e^{-iE_0 t} \, \Gamma \, {\left | \psi \right \rangle}_i,
\end{equation}
where the exponential term corresponds to the dynamical phase (which can be set to zero by taking \(E_0 = 0\)) and \(\Gamma = e^{i\gamma}\) represents the geometric phase operator. The parameter \(\gamma\) depends only on the geometry or topology of the trajectory in parameter space and is independent of how quickly the evolution is performed, provided adiabaticity is preserved. For graphene-based nanostructures, geometric phases can emerge due to lattice deformations or the presence of topological defects such as pentagons or heptagons. These defects break the perfect symmetry of the hexagonal lattice, inducing effective gauge fields that mimic a magnetic flux. This fictitious flux gives rise to a Berry-type phase that can be computed in the following form,
\begin{equation}
\gamma = \iint_S \mathbf{B} \cdot d\mathbf{S},
\end{equation}
where \(\mathbf{B}\) is the emergent magnetic field associated with the defect-induced curvature. 

\begin{figure}[t] 
    \centering
    \includegraphics[width=0.5\linewidth]{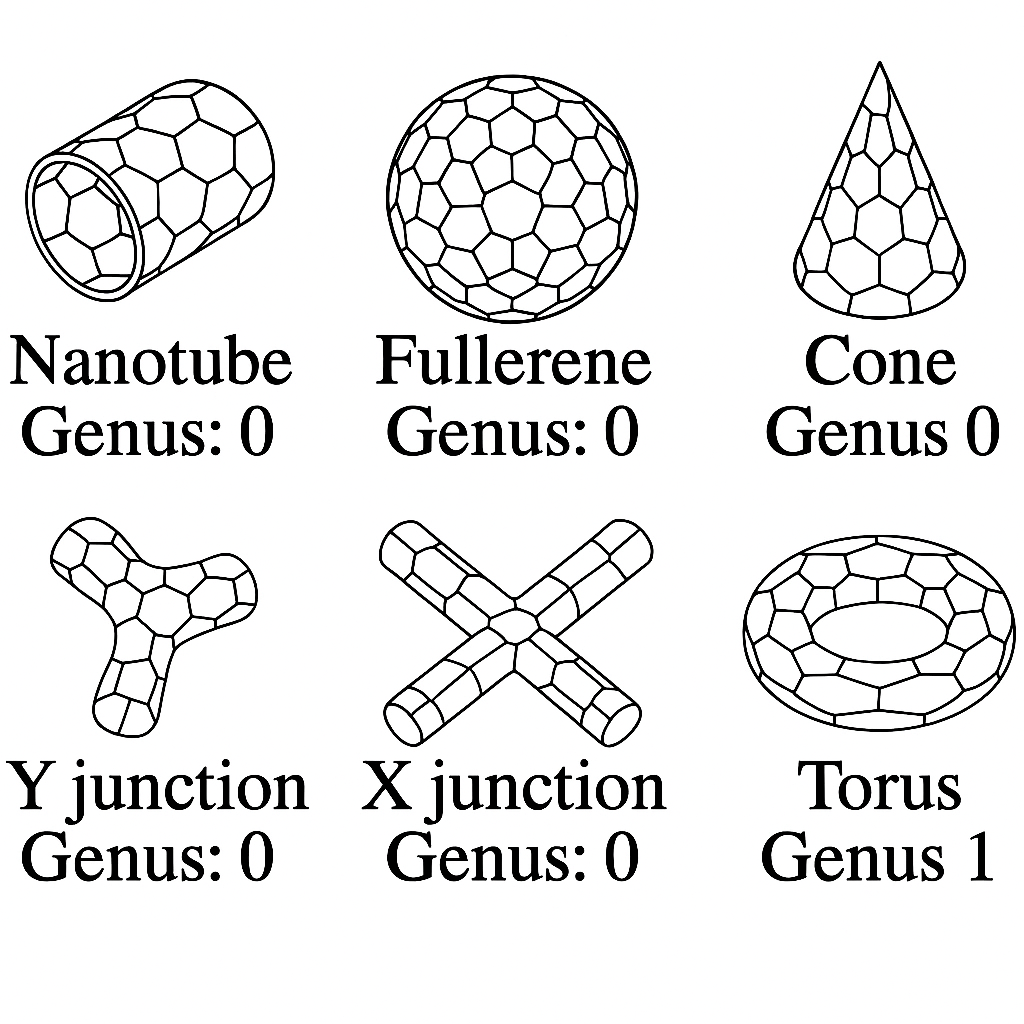}
    \caption{Graphene-derived nanostructures with different topologies. Each object is labeled with its genus $g$, which counts the number of holes or handles and determines the global topological class of the surface.}
    \label{fig:genus_structures}
\end{figure}

As discussed in Eq.~(\ref{index}), the geometric phase can be directly connected to the topological characteristics of the structure, specifically its genus \(g\) and the number of open faces \(N\). The resulting phase is given by:
\begin{equation}\label{gual}
\gamma = 3\pi \left[(1 - g) - \frac{N}{2}\right].
\end{equation}
Equation~(\ref{gual}) provides a simple and powerful way to compute the geometric phase for each spinor component, from purely topological data, without solving the full quantum dynamics. Table~\ref{tab1} lists the values of \(\gamma\) for several nanostructures experimentally realized from graphene. These include closed and open systems with different values of genus and boundary count. Although higher-genus structures with octagonal rings have been proposed and shown to be stable~\cite{PhysRevB.55.15.1997}, here we focus on those built from pentagonal and hexagonal cells. Fig.~\ref{fig:genus_structures} visually complements table~\ref{tab1} by displaying the topological classes of typical graphene-based nanostructures. The genus \( g \) distinguishes between structures with or without handles: while the fullerene, cone, nanotube, and junctions all share \( g = 0 \), the torus features a single handle, and thus \( g = 1 \). This topological distinction plays a direct role in determining the total geometric phase acquired by a spinor during parallel transport, as encoded in Eq.~(\ref{gual}). This leads to a quantized holonomy for spinorial wavefunctions under parallel transport along a closed path:
\begin{eqnarray}
    U(n) &=& exp\left(in\pi\tau_y\right);\nonumber\\
         &=& \cos(n\pi) + i\tau_y \sin{(n\pi)},
\end{eqnarray}
or in matrix form,
\begin{equation}
    U(n) = \begin{pmatrix}
        \cos(n\pi) & \sin(n\pi)\\
        -\sin(n\pi) & \cos(n\pi)
    \end{pmatrix}.\label{MatrixHolo}
\end{equation}
Specifically, after a rotation of \(2\pi\) and for values of $n$ integer, the spinor transforms as
\begin{equation}
\Psi \rightarrow e^{in\pi} \Psi = \pm \Psi,
\end{equation}
depending on whether \(n\) is even or odd. Hence, the acquired geometric phase determines whether the spinor returns to itself or changes sign, which in turn affects the interference and coherence properties in quantum transport phenomena. This behavior, like \(\mathbb{Z}_2\), emerges from the global topology of the nanostructure and exemplifies how geometry and topology can be encoded in the quantum phase of a system. Such effects are not merely theoretical; they can influence observable quantities such as conductance or the robustness of quantum gates, making them relevant for quantum technologies based on graphene and other two-dimensional materials.

\begin{table}[t]
\centering
\caption{Geometric phases of nanostructures derived from graphene.}
\label{tab1}
\begin{tabular}{cccc}
\hline
Structure      & Genus ($g$) & Open faces ($N$) & Phase ($\gamma$) \\
\hline
Nanotube       & 0           & 2                & $0$              \\
Fullerene      & 0           & 0                & $3\pi$           \\
Cone           & 0           & 1                & $\frac{3\pi}{2}$           \\
Y junction     & 0           & 3                & $-\frac{3\pi}{2}$          \\
X junction     & 0           & 4                & $-3\pi$          \\
Torus          & 1           & 0                & $0$              \\
\hline \\
\end{tabular}
\end{table}

\section{Holonomic logical gates}

From the Atiyah-Singer geometric phase, we have built a one-qubit 
on graphene based on its valley degree of freedom. With this purpose, we have to consider the holonomy~\eqref{GeoHolo} as a unitary matrix. Let us define a quantum logical base for qubits, with a focus on holonomy~\eqref{MatrixHolo}, as follows:
\begin{subequations}
\begin{eqnarray}
    |0\rangle &=& |K, A/B\rangle;\\
    |1\rangle &=& |K', A/B\rangle,
\end{eqnarray}
\end{subequations}
where $\tau_i$ acts on the valley degree of freedom, causing a mixing between $K$ and $K'$ Fermi points. Since the holonomy~\eqref{GeoHolo} depends on the $\sigma_z$ Pauli matrix, it does not interfere with holonomy~\eqref{MatrixHolo}, and we can treat it independently.

Our main result lies in the analysis of nanostructures with geometric phase as multiples of $\pi/2$, as occurs for the nanocone and Y junction (see Table~\ref{tab1}). For these cases, we can build a Pauli-Y gate~\cite{Nielsen} from holonomy~\eqref{MatrixHolo}, 
\begin{equation}
    \mathcal{Q}_Y = U\left(\pm \tfrac{3\pi}{2}\right) = e^{\pm i\frac{3\pi}{2}} \begin{pmatrix}
        0 & -i\\
        i & 0
    \end{pmatrix},\label{QY-Gate}
\end{equation}
where \( n = 3(1-g) - \frac{3N}{2} \), and the `plus' sign is related to nanocones and the `minus' sign is related to Y junctions. Such a Pauli-Y gate is responsible for a phase change of the Fermi points on the qubit basis. Note that, if we work with these two holonomies \eqref{GeoHolo} and \eqref{MatrixHolo} together, we are capable of building more complex quantum holonomic gates in order to perform holonomic quantum computation.

\section{Conclusions}

In this work, we presented a topological approach to calculate the geometric phase acquired by graphene-based molecules using only their global characteristics, such as the genus \( g \) and the number of open faces \( N \). Based on the Atiyah-Singer index theorem~\cite{AnnMath.87.3.1968}, we related the number of zero modes in the Dirac operator to an effective gauge flux induced by topological defects, such as pentagonal and heptagonal rings inserted into the hexagonal graphene lattice. We observed that, when the geometric phase \(\gamma\) acquired by a spinor under parallel transport is quantized in integer multiples of \(\pi\) leads to a binary behavior under full $2\pi$ rotations: when \( |n| \) is even, the spinor returns to its original configuration (\( \Psi \to \Psi \)), while for odd \( |n| \), a sign change occurs (\( \Psi \to -\Psi \)). For half-integer values of $n$, the holonomy~\eqref{MatrixHolo} can be described as a Pauli-Y gate. This type of quantum logic gate performs a phase change in combination with a state change on the valley degree of freedom of the logical basis. And the possibility of dealing with both holonomies~\eqref{GeoHolo} and~\eqref{MatrixHolo} gives us a way to build more complex quantum logical gates for graphene.

This distinction plays a fundamental role in the interference properties of quantum states confined to curved graphene geometries. Such geometric phases influence the dynamics and possible interference patterns between quasiparticles in closed circuits over these nanostructures. Therefore, constructive or destructive interference between quantum states is not determined by local fields but by the global topology of the molecule, highlighting the geometric and topological nature of the phase. Importantly, the theoretical predictions derived here are experimentally accessible. As shown in~\cite{Nature.388.6641.1997}, techniques such as hydrocarbon pyrolysis in carbon arcs can generate a variety of curved graphene-like structures, such as fullerenes, nanotubes, and conical junctions, each characterized by different values of genus and boundary number. These systems provide natural laboratories for testing the validity of the proposed framework and for exploring the role of topological phases in quantum transport. The results obtained reinforce the idea that the topological structure of carbon nanostructures directly determines their quantum properties. This provides a bridge between discrete lattice models and continuum geometric field theories and opens promising avenues for the development of holonomic quantum computation based on graphene nanostructures.

\appendix
\section{Atiyah-Singer Index Theorem}\label{appxA}

In order to obtain a straight relation between the graphene zero modes and the total flux from the curvature and the topological defect, we have made use of the Atiyah-Singer index theorem. Therefore, let us rewrite the Dirac operator of \eqref{curvatura} in terms of new operators $P$ and $P^{\dagger}$, as expressed by Pachos in ref.~\cite{ContempPhys.50.2.2009},
\begin{equation}\label{operador}
D\!\!\!\!/ = \begin{pmatrix} 0 & P^\dagger \\ P & 0 \end{pmatrix},
\end{equation}
where \( P \) maps the eigenstates of the subspace \( V^+ \) to \( V^- \), and \( P^\dagger \) maps from \( V^- \) back to \( V^+ \). In other words, we are interested in the solutions of the Dirac equation where \( D\!\!\!\!/ \Psi = 0 \). Let \( \nu_+ \) be the number of zero-energy eigenstates associated with the operator \( P^\dagger P \), and \( \nu_- \) the number associated with \( P P^\dagger \). To analyze the spectral asymmetry, we have introduced the chirality operator as given below. Thus,
\begin{equation}\label{quiralidade}
\gamma_5 = \begin{pmatrix} 1 & 0 \\ 0 & -1 \end{pmatrix}.
\end{equation}
And, instead of working with \( D\!\!\!\!/ \) directly, we have using \( D\!\!\!\!/^2 \) operator, since it shares the same spectrum (eigenvalue count) and is easier to handle due to its diagonal structure. Then 
\begin{equation}\label{quadrado}
D\!\!\!\!/^2 = \begin{pmatrix} P^\dagger P & 0 \\ 0 & P P^\dagger \end{pmatrix}.
\end{equation}
Note that, once the operators \( P^\dagger P \) and \( P P^\dagger \) present the same $\lambda$ nonzero eigenvalues corresponding to the $P^{\dagger}\psi$ eigenstate, i.e., $PP^{\dagger}\psi = P^{\dagger} P\psi = \lambda\psi$, this means that the nonzero parts of the spectra of \( P^\dagger P \) and \( P P^\dagger \) coincide, and they share the $\lambda$ eigenvalue with the operator. 

Nevertheless, this is not guaranteed for zero eigenvalues: the dimensions of the null spaces of these operators can differ. This asymmetry is at the heart of the index theorem. To proceed, we consider the trace expression as
\begin{equation}\label{expansao}
\operatorname{Tr}(\hat{f} e^{-t \hat{D}}) = \frac{1}{4 \pi t} \sum_{k \geq 0} t^{k/2} a_k(\hat{f}, \hat{D}),
\end{equation}
known as the heat kernel expansion, where \( a_k \) are the heat kernel coefficients. Setting \( \hat{f} = \gamma_5 \) and \( \hat{D} = D\!\!\!\!/^2 \), and recalling that the trace of a commutator vanishes for nonzero eigenvalues, we obtain
\begin{equation}\label{auto}
\operatorname{Tr}(\gamma_5 e^{-t D\!\!\!\!/^2}) = \sum_{\lambda_+} e^{-t \lambda_+} - \sum_{\lambda_-} e^{-t \lambda_-},
\end{equation}
where \( \lambda_+ \) and \( \lambda_- \) are eigenvalues of \( P^\dagger P \) and \( P P^\dagger \), respectively. Since the nonzero parts cancel, the trace reduces to the difference in the number of zero modes given by
\begin{equation}\label{numero}
\operatorname{Tr}(\gamma_5 e^{-t D\!\!\!\!/^2}) = \nu_+ - \nu_-.
\end{equation}

From the heat kernel expansion, only the term independent of \( t \) survives in the trace involving \( \gamma_5 \). Therefore,
\begin{equation}\label{result}
\operatorname{Tr}(\gamma_5 e^{-t D\!\!\!\!/^2}) = \frac{a_2}{4 \pi}.
\end{equation}
The coefficient \( a_2 \) is well known in the literature and can be computed from the operator:
\begin{equation}
D\!\!\!\!/^2 = -g^{\mu \nu} \nabla_\mu \nabla_\nu + \frac{i}{4} [\gamma^\mu, \gamma^\nu] F_{\mu \nu} - \frac{1}{4} R,
\end{equation}
where \( F_{\mu \nu} = \partial_\mu A_\nu - \partial_\nu A_\mu \) and \( R \) is the scalar curvature. The magnetic field is given by \( B_k = \frac{1}{2} \epsilon^{k \mu \nu} F_{\mu \nu} \), and the coefficient becomes,
\begin{equation}\label{a2}
a_2 = \operatorname{Tr} \left\lbrace \gamma_5 \left( \frac{i}{4} [\gamma^\mu, \gamma^\nu] F_{\mu \nu} - \frac{1}{4} R \right) \right\rbrace = 2 \iint \mathbf{B} \cdot d\mathbf{S}.
\end{equation}
Combining with Eq.~\eqref{numero}, we finally arrive at the Atiyah-Singer index Theorem in this context. We have
\begin{equation}\label{final}
\text{ind}(D\!\!\!\!/) = \nu_+ - \nu_- = \frac{1}{2\pi} \iint \mathbf{B} \cdot d\mathbf{S}.
\end{equation}
This result links the difference in the number of zero-energy eigenstates to the total magnetic flux through the surface. In the case of graphene and its geometrical variants, this flux can often be computed via the Gauss-Bonnet theorem, providing a direct bridge between topology and electronic properties.

\section*{Declaration of Interests}

The authors declare that they have no known competing financial interests 
or personal relationships that could have appeared to influence the work 
reported in this paper.\\

{\bf Acknowledgements:} This work was supported by Conselho Nacional de Desenvolvimento Cient\'{\i}fico e Tecnol\'{o}gico (CNPq) and Funda\c{c}\~ao de Apoio a Pesquisa do Estado da Para\'iba (Fapesq-PB). G. Q. Garcia would like to thank Fapesq-PB for financial support (Grant BLD-ADT-A2377/2024). The work by C. Furtado is supported by the CNPq (project PQ Grant 1A No. 311781/2021-7).

 
\end{document}